\documentclass[aps,prd,reprint,superscriptaddress,showpacs]{revtex4-1}
\usepackage{graphicx}
\usepackage{amsmath,amssymb}
\usepackage{hyperref}
\pdfoutput=1

\begin{document}
\title{Classifying Pole of Amplitude Using Deep Neural Network}
\author{Denny Lane B. Sombillo}
\email[]{sombillo@rcnp.osaka-u.ac.jp}	
\affiliation{National Institute of Physics, University of the Philippines Diliman, Quezon City 1101, Philippines}	
\affiliation{Research Center for Nuclear Physics (RCNP), Osaka University, Osaka 567-0047, Japan}
		
\author{Yoichi Ikeda}
\affiliation{Kyushu University, Fukuoka 819-0395, Japan}
\author{Toru Sato}
\author{Atsushi Hosaka}
\affiliation{Research Center for Nuclear Physics (RCNP), Osaka University, Osaka 567-0047, Japan}

\begin{abstract}
Most of exotic resonances observed in the past decade appear as peak structure near some threshold. These near-threshold phenomena can be interpreted as genuine resonant states or enhanced threshold cusps. Apparently, there is no straightforward way of distinguishing the two structures. In this work, we employ the strength of deep feed-forward neural network in classifying objects with almost similar features. We construct a neural network model with scattering amplitude as input and nature of pole causing the enhancement as output. The training data is generated by an S-matrix satisfying the unitarity and analyticity requirements. Using the separable potential model, we generate a validation data set to measure the network's predictive power. We find that our trained neural network model gives high accuracy when the cut-off parameter of the validation data is within $400$-$800\mbox{ MeV}$. As a final test, we use the Nijmegen partial wave and potential models for nucleon-nucleon scattering and show that the network gives the correct nature of pole.
\end{abstract}		
\maketitle
\section{Introduction}
\label{sec:0}
Renewed interest in hadron spectroscopy started after the discovery of $X(3872)$ in 2003 \cite{Choi2003}. Since then, several candidates of nonstandard exotic hadrons are proposed. One common feature of these phenomena is that they manifest as sharp peak structure near some threshold \cite{Olsen2018}. The proximity of an enhancement to the threshold introduces several possible nature of peak's origin. One of the appealing possibilities is a weakly bounded hadronic molecule composed of two hadrons \cite{Guo2018,Yamaguchi2020} which can be associated to the presence of a pole near the two-particle threshold. Other possibilities are  purely kinematical in nature such as cusps or triangle singularities \cite{CuspTriangle2020}. On one hand, threshold cusp is always present in s-wave scattering whenever an inelastic channel opens. However, it has been shown in  \cite{Frazer1964,PearceGibson,Cusp2015,Swanson2016} that threshold cusp can only produce a significant enhancement provided that there is some near-threshold pole even if it is not located in the relevant region of unphysical sheet. On the other hand, triangle singularity does not need nearby pole to produce a pronounced enhancement but instead requires that intermediate particles be simultaneously on-shell \cite{CuspTriangle2020,Landau1959,Nakamura2019,Nakamura2019p2}. 
	
The purpose of this paper is to address the origin of sharp peak observed around the threshold of two-body hadron scattering problems. We specifically focus on the case where a near-threshold pole causes the peak structure and attempt to identify its nature, i.e. whether it is bound, resonance or virtual state pole. Until now, there has not been a method to distinguish the pole origin of peak structure around the threshold. In general, this is a difficult program because of the limited resolution of experimental data. 
	
Here, we treat the identification of the nature of pole causing the enhancement as a classification task \cite{Alpaydin2010} and solve it using supervised machine learning. The machine learning technique is ubiquitous even in physical sciences \cite{MLPhys} and it is well known that deep neural network excel in solving a classification task. In this work we demonstrate how a deep neural network can be applied to identify the pole origin of cross-section enhancement.  This includes defining the appropriate input-output data, setting up the network architecture and generating the training dataset. As a first effort to apply deep learning in the classification of pole causing a cross-section enhancement, we only consider here the single-channel scattering.

This paper is organized as follows. In section \ref{sec:1} we give a short background on how a neural network works. One of the crucial part of deep learning is the preparation of dataset. In section \ref{sec:2} we describe how the training dataset is generated using the general properties of S-matrix. The performance of our neural network model using the training dataset is discussed in section \ref{sec:3}. In section \ref{sec:4} we explore the applicability of our trained network using a separable potential model to generate a validation dataset. We also use the partial wave and potential models of Nijmegen group as a final test in the same section. Finally, we formulate our conclusion in section \ref{sec:6}.

\begin{figure*}[t!]
	\centering
	\includegraphics[width=0.9\linewidth]{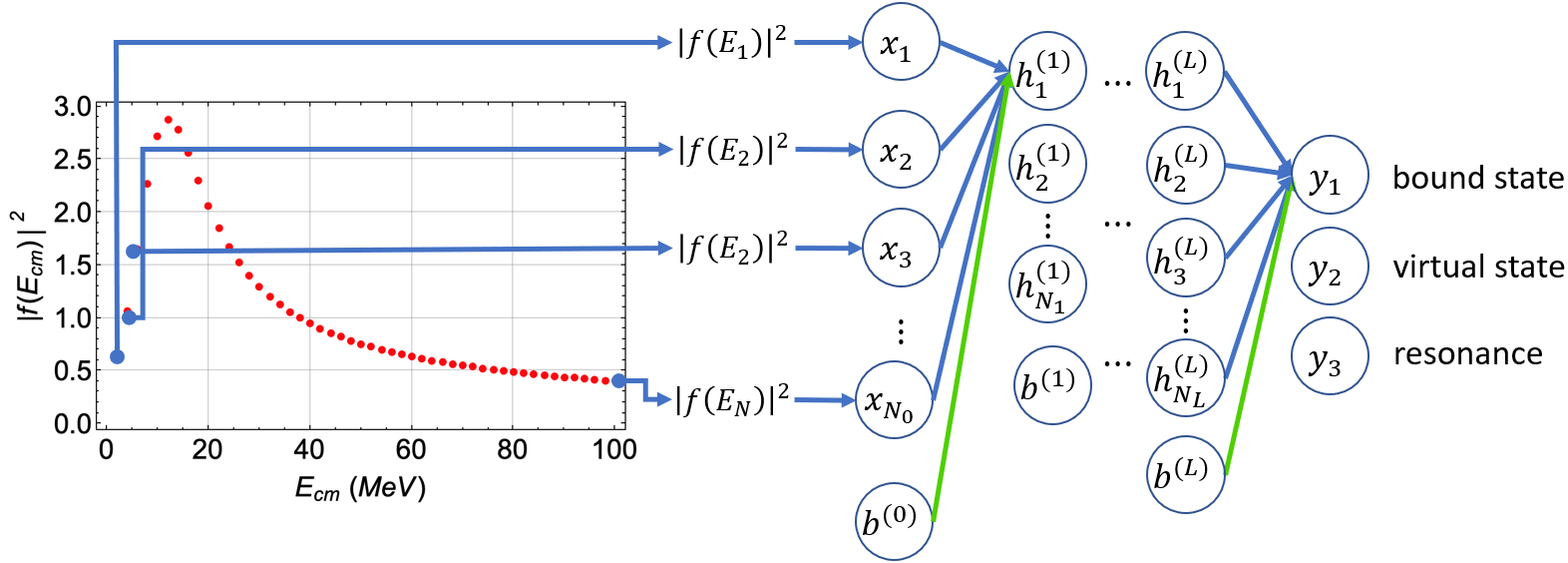}
	\caption{Schematic of deep neural network for S-matrix pole classification.}
	\label{fig:network}
\end{figure*}		
\section{Deep Neural Network for Pole Classification}
\label{sec:1}
We briefly review the basic operation in deep learning \cite{Nielsen} and discuss how it can be applied to pole classification problem. A neural network consists of an input, hidden layers and an output layer where each layer contains certain number of nodes. We use the term deep neural network for architectures having more than one hidden layers. Fig. \ref{fig:network} shows the deep neural network set-up that we used in this study. The nodes $x_i$'s in the input layer contain numerical values describing certain features of the input data while nodes that are not in the input layer are equipped with activation functions with range $(0,1)$ or $(0,\infty)$, whichever is applicable. The nodes in $(\ell-1)^{th}$ layer are sent to each $\ell^{th}$ layer node by putting them in a linear combination
\begin{equation}
	z^{(\ell)}_i=\Sigma_{j} w^{(\ell)}_{ij}h^{(\ell-1)}_j+b^{(\ell-1)}
	\label{eq:preactivation}
\end{equation}
where $z^{(\ell)}_i$ is the $i^{th}$ node pre-activation value in the $\ell^{th}$ layer , $h^{(\ell-1)}_j$ is the $j^{th}$ node post-activation value of the $(\ell-1)^{th}$ layer, $w^{\ell}_{ij}$ is the weight connecting $j^{th}$ node of $(\ell-1)^{th}$ layer to $i^{th}$ node of $\ell^{th}$ layer and $b^{(\ell-1)}$ is the bias in $(\ell-1)^{th}$ layer. In this notation, input nodes are represented as $x_i=h^{(0)}_i$. The pre-activation value $z^{(\ell)}_i$ is fed to the activation function to get the node's post-activation value:
\begin{equation}
	h^{(\ell)}_i=\sigma(z^{(\ell)}_i).
	\label{eq:activation}
\end{equation}
This arrangement of layers and nodes together with the choice of activation functions allows the neural network to build a nonlinear mapping of input vector $\mathbf{x}$ to output vector $\mathbf{y}$.

The goal of deep learning is to find an optimal mapping between $\mathbf{x}$ and $\mathbf{y}$. To do this, one has to prepare a training dataset containing inputs  with known outputs. Initially, some random weights and biases are assigned to the neural network. Then we perform a forward pass, i.e. we feed all the training inputs and let the network calculate all the outputs. Now, the average difference between true output and network's output define the cost function $C(\hat{w},\vec{b})$ where $\hat{w}$ and $\vec{b}$ are the initial weight matrix and bias vector, respectively. The weights and biases are updated using the gradient descent method via backpropagation \cite{Backprop1986}. One forward pass together with one backpropagation of the entire training dataset comprise one epoch of training. Several epochs are normally executed to update the weights and biases until the cost function reached its global minimum. The neural network architecture with its updated weights and biases correspond to the optimal map that we seek. 
	
In this study, we construct a deep neural network with the cross-section of two-body scattering, $|f(E_{cm})|^2$, on a discretized center-of-mass energy axis $[0,100\mbox{ MeV}]$ with $0.5\mbox{ MeV}$ spacing as input and a vector with three elements as output. One can use smaller spacing but this will increase the number of input nodes and may result into slow cost function convergence. Similarly, taking larger spacing with smaller number of input nodes will most likely converge fast to a higher cost function minimum. The chosen 0.5 MeV provides an optimal spacing for the current study. Now, the output nodes correspond to three distinct pole classifications, i.e. bound state, virtual state or resonance as shown in Fig.\ref{fig:network}. The classification of pole is described as follows. Suppose $p_0$ represents the pole position on the complex momentum plane $\mathbb{C}$, then we say that it is a bound state pole if $p_0$ is positive pure imaginary. If $\mbox{Im }p_0<0$ and $|\mbox{Im }p_0|>|\mbox{Re }p_0|$, then $p_0$ is a virtual state pole. Otherwise, if $|\mbox{Im }p_0|<|\mbox{Re }p_0|$ we call it resonance (see Appendix of \cite{Virtual2011} for detailed explanation). 
	
To obtain the optimal values of weights and biases, the network must be trained using a dataset of cross-section with known enhancement origin. This will be explained in the next section.
	
\section{Dataset to train deep neural network}
\label{sec:2}
\subsection{General Properties of S-Matrix}
Ideally, a reliable neural network model that can distinguish the nature of pole responsible to the cross-section enhancement must be optimized using a training dataset generated from an exact S-matrix. However, such an S-matrix cannot be derived from the fundamental theory of strong interaction QCD for hadrons due to its non-perturbative nature. In such a situation, we can still deduce the general form of S-matrix using the analyticity and unitarity requirements \cite{Chew1,Chew2, AnalyticSMatrix1966}. 

\begin{figure}[t!]
	\centering
	\includegraphics[width=0.6\linewidth]{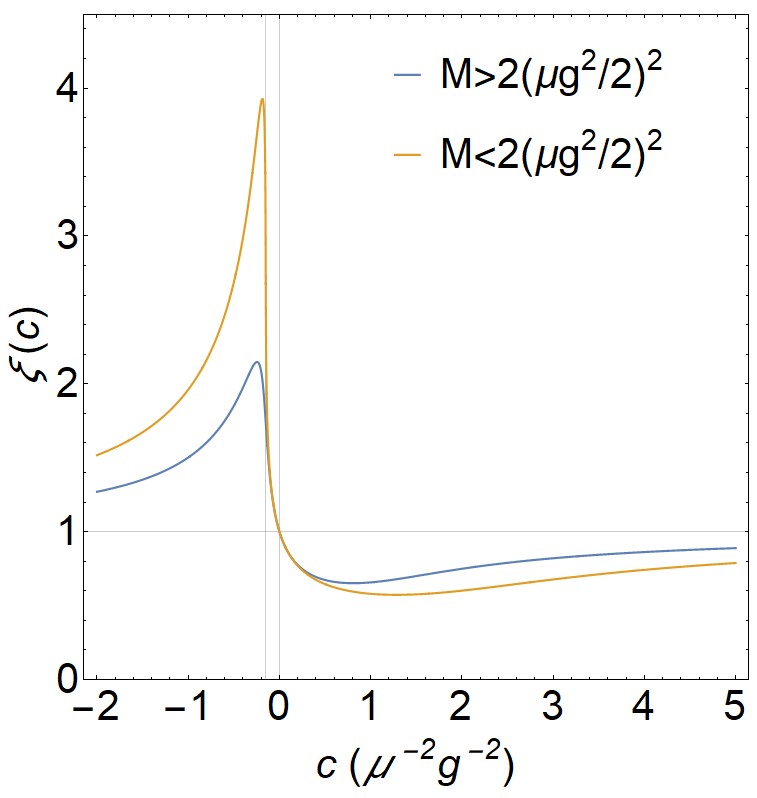}
	\caption{Behavior of $\xi=\xi(c)$.}
	\label{fig:xi}
\end{figure}	   	
	
Consider the s-wave scattering of two particles with mass $m_1$ and $m_2$, reduced mass $\mu=m_1m_2/(m_1+m_2)$ and relative momentum magnitude of $p$. The S-matrix can be parametrized as 
\begin{equation}
	S(p)=\frac{1-i\mu p K(p)}{1+i\mu p K(p)}
	\label{eq:smatrix1}
\end{equation}
satisfying unitarity provided that  $K(p)$ is the real-valued K-matrix \cite{Taylor, Newton, Suzuki}. At energies near the location of K-matrix pole $M'$, we can write $K=g'^2/(E-M')+c$ where $E=E_1+E_2$ with $E_i$ as the energy of particle $m_i$ and $g',c$ are reals. Analyticity and $K(-p)=K(p)$ are satisfied in the non-relativistic case, i.e. $E=p^2/(2\mu)$,  by the parametrization
\begin{equation}
	K(p)=\frac{g^2}{p^2-M}+c
	\label{eq:kmatrix}
\end{equation} 
where $g^2=2\mu g'^2$ and $M=2\mu(M'-m_1-m_2)$. From the S-matrix  in \eqref{eq:smatrix1}, one can obtain the partial wave amplitude using the relation 
\begin{equation}
	S(p)=1+2ipf(p)
	\label{eq:pwa}
\end{equation}

\begin{figure*}[t!]
	\centering
	\includegraphics[width=0.3\textwidth]
	{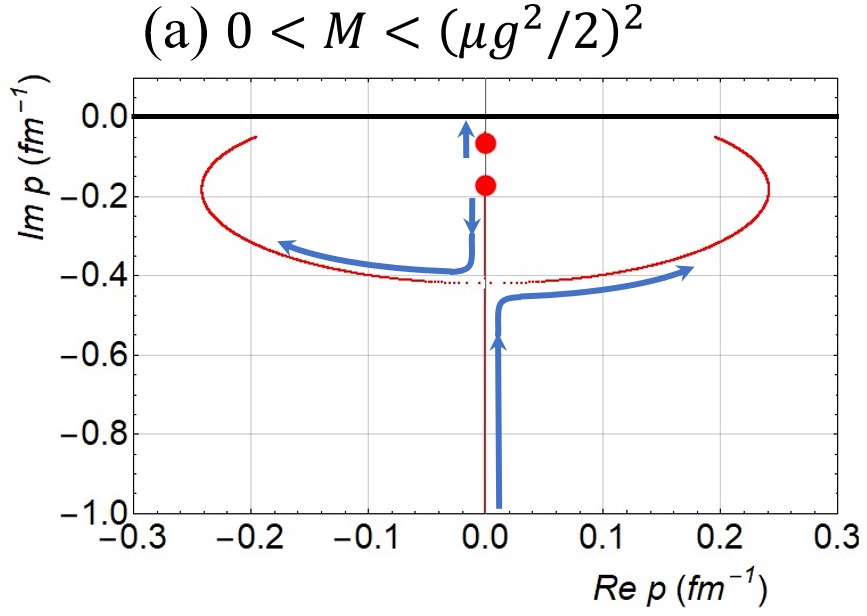}
	\includegraphics[width=0.3\textwidth]
	{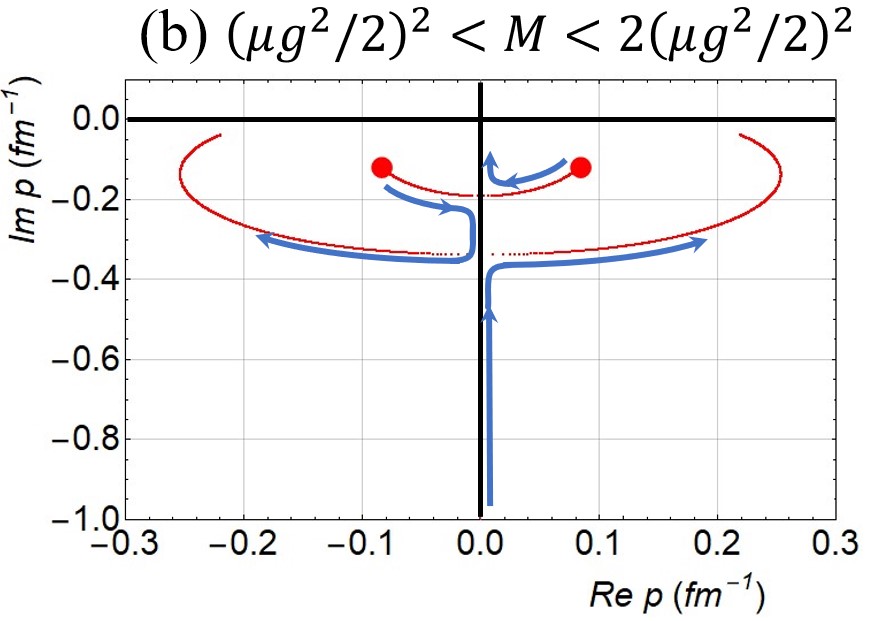}
	\includegraphics[width=0.3\textwidth]
	{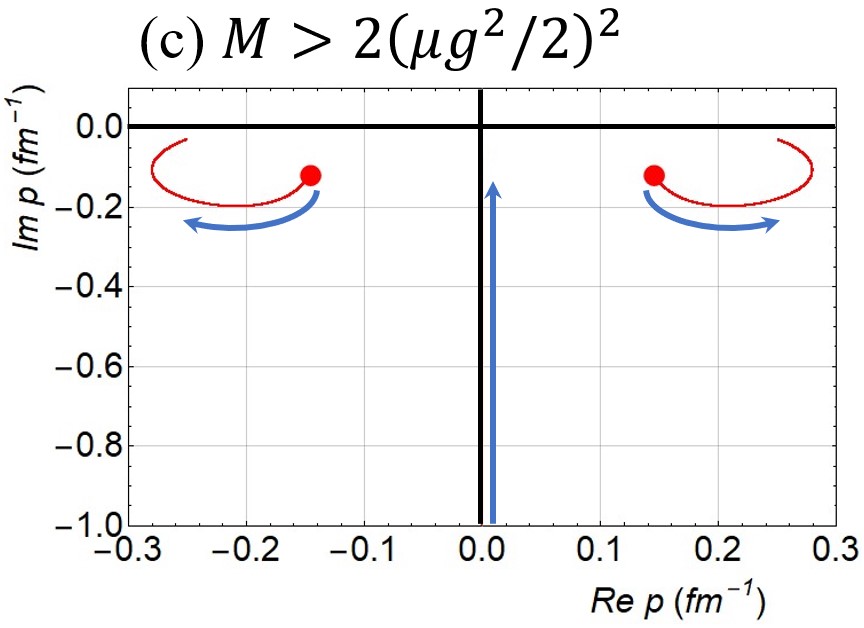}
	\caption{Configuration of S-matrix pole as $c$ is varied from zero to some negative value starting with (a) a pair of virtual states, (b) virtual state with widths and (c) resonance. The red dots represents the pole positions when $c=0$, the red line shows the trajectory and the blue line shows the direction of pole motion as $c$ becomes negative.}
	\label{fig:KMatrixczero}
\end{figure*}

Consider now how the K-matrix parameters dictate the singularities of S-matrix in \eqref{eq:smatrix1}. If we substitute $K(p)$ into $S(p)$, we get
\begin{equation}
	S(p)=-\frac{i\mu c p^3-p^2+i\mu(g^2-cM)p +M}
	{i\mu c p^3+p^2+i\mu(g^2-cM)p -M}
	\label{eq:smatrix}
\end{equation}	
and the pole position is obtained from
\begin{equation}
	i\mu c p^3+p^2+i\mu(g^2-cM)p -M = 0.
	\label{eq:polecubic}
\end{equation}
Taking the complex conjugate of \eqref{eq:polecubic} and knowing that $\mu, g^2$ and $M$ are reals, we can recover the same equation as that for $p$, i.e. $-p^*$ satisfies the same cubic equation. This means that the denominator of $S(p)$ in \eqref{eq:smatrix} contains a factor $(p+i\beta)^2-\alpha^2$ which gives a conjugate pair of poles with real $\alpha, \beta$. The third unpaired solution to \eqref{eq:polecubic} must have the property $p=-p^*$. This can only be true if $p$ is pure imaginary. In fact it is possible that all the solutions of \eqref{eq:polecubic} are pure imaginary. It follows that we can write \eqref{eq:smatrix} in factored form as
\begin{equation}
	S(p)=\left(-\frac{p+i\gamma}{p-i\gamma}\right)
	\left[\frac{(p-i\beta)^2-\alpha^2}{(p+i\beta)^2-\alpha^2}\right]
	\label{eq:factored}
\end{equation}
 where $\alpha,\beta,\gamma$ are real numbers that are related to $g^2$, $M$ and $c$ parameters. 
	 	 	
For $c=0$ we only have a pair of conjugate poles given by
\begin{equation}
	p_0 = -i\frac{\mu g^2}{2}\pm\sqrt{M-\left(\frac{\mu g^2}{2}\right)^2}
	\label{eq:czero}
\end{equation}
and we readily identified $\beta=\mu g^2/2$ and $\alpha=\sqrt{M-(\mu g^2/2)^2}$. 
Note that $\beta>0$ is required to avoid having S-matrix poles on the upper half momentum plane (other than the imaginary axis), otherwise causality is violated \cite{AnalyticSMatrix1966,Kampen1953}. For $c\neq 0$, a third imaginary pole $i\gamma$ appears and $\alpha,\beta$ are modified according to:
\begin{equation}
	\begin{split}
	\alpha^2  &=\xi M - \beta^2 \\
	\beta &= \frac{\mu g^2}{2}\xi\left[1+(1-\xi)\dfrac{Mc}{g^2}\right] \\
	\gamma &=\frac{1}{\xi \mu c}.
	\end{split}
	\label{eq:cnotzero}
\end{equation}
These are obtained by comparing the expansion in the denominator of \eqref{eq:factored} with that of \eqref{eq:polecubic}. A dimensionless quantity $\xi$ is introduced to facilitate the comparison and for given values of $\mu, g^2$ and $M$, $\xi$ is an implicit function of $c$ given by
\begin{equation}
	1=\xi+c\mu^2g^2\xi^2\left[1-(1-\xi)\dfrac{cM}{g^2}\right]
	\label{eq:xicubic}
\end{equation}
with $\xi\rightarrow1$ as $c\rightarrow0$ or $c\rightarrow\pm\infty$ (see Fig.\ref{fig:xi}). 

The bounded $\xi=\xi(c)$ implies that the third pole $i\gamma$ will originate from $\pm\infty i$ as $c$ becomes nonzero. For $c>0$, we can generate a simple pole at $p_0=i\gamma$ on the upper half momentum plane and if we let $c\rightarrow+\infty$, this pole gets very close to the threshold. This corresponds to a bound state in accordance to the completeness relation in \cite{NingHu1948}. Now, as we vary $c$ from zero to some negative value, the poles redistribute themselves as shown in Fig.\ref{fig:KMatrixczero}. Here, we see an instance when all the three poles are pure imaginary and at some finite values of $c$, two of the poles will merge and turn into conjugate pair as seen in Fig.\ref{fig:KMatrixczero}(a) and (b). The merging of poles happens at some $c<0$ when the slope of $\xi$ becomes infinite as shown in Fig.\ref{fig:xi}.  This demonstrates that the constant term in \eqref{eq:kmatrix} is capable of generating S-matrix pole and should not be treated as background (see also \cite{Workman2012}). 
 	
The conjugate pair of poles in \eqref{eq:factored} will always have $\beta>0$ for all values of $c$. For $c\rightarrow 0$, $\xi$ approaches unity and we recover \eqref{eq:czero} with $\beta>0$. Also, as $c\rightarrow+\infty$, \eqref{eq:cnotzero} gives a positive $\beta$  since $0<\xi<1$. Finally, if $c<0$ we see from Fig.\ref{fig:xi} that $(1-\xi)<0$ and this still gives a positive $\beta$ demonstrating that causality is not violated for all values of $c$.  	
 	
The form of S-matrix in \eqref{eq:factored} and its relation to K-matrix in \eqref{eq:kmatrix} allows us to identify the parenthetical factor as the generator of pure imaginary momentum pole and the square-bracket factor as the generator of conjugate poles. To avoid ambiguity in the classification it is more plausible to separate these two factors. That is, the first factor will only be used to generate the bound-virtual dataset while the second factor will be used to generate conjugate virtual-resonance dataset. The two datasets will be combined as a single classification dataset before we use it to optimize the parameters of our neural network. This will suffice to assign three distinct outputs in our neural network, i.e. bound, virtual and resonance. Note that one can also use the combined form in \eqref{eq:factored} but a ``bound with resonance" must be added to the output entry. This additional category is not yet relevant in the current study.
	
\subsection{bound state and virtual state}
Let us first consider the threshold-enhancement caused by a shallow bound state or a virtual state in s-wave amplitude. From previous discussion, we learned that the first factor in \eqref{eq:factored} can be used to generate a near-threshold bound or virtual state pole. A closer look will reveal that this gives an identical cross-section whatever the sign of $\gamma$. That is, with $S(p)=-(p+i\gamma)/(p-i\gamma)$ we get $|f(p)|^2=1/(p^2+\gamma^2)$ and there is no way to distinguish between virtual and bound state enhancements. In general, there is background contribution in addition to the pole part of S-matrix, making it possible to distinguish the two enhancements. Thus, it is imperative to include a background to the S-matrix parametrization for the bound-virtual dataset, i.e. 
\begin{equation}
	S(p)=e^{2i\delta_{bg}(p)}
	\left(-\dfrac{p+i\gamma}{p-i\gamma}\right).
	\label{eq:sforbound}
\end{equation}
where $\delta_{bg}(p)$ is the background phase.	
	
The form of $\delta_{bg}(p)$ is restricted by unitarity and analyticity requirements. First, unitarity dictates that $\delta_{bg}(p)$ be a real-valued function for real momentum $p$. Second, analyticity requires that there be no poles in the analytically continued $e^{2i\delta_{bg}(p)}$ on the upper-half momentum plane and that the reflection principle be satisfied.  Here, we introduce the background phase shift given by
\begin{equation}
	\delta_{bg}(p)=\eta \tan^{-1}\left(\dfrac{p}{\Lambda_{bg}}\right).
	\label{eq:bg}
\end{equation} 
where $\eta$ is a real parameter and $\Lambda_{bg}>0$ is the training S-matrix cut-off parameter. If we let $\eta<0$, \eqref{eq:bg} reduces to a repulsive hard-core type  background used in \cite{IshidaBG1997} with $-\eta/\Lambda_{bg}$ as the core radius if $p$ is near the threshold. Also, \eqref{eq:bg} can simulate the left-hand cut both in the physical and unphysical sheet even in the non-relativistic case since the analytically continued $\tan^{-1}(p/\Lambda_{bg})$ has branch cuts in $\mathbb{C}$ along the strip $(-i\infty,-i\Lambda_{bg})\cup(i\Lambda_{bg},i\infty)$\cite{Gamelin}.
	
Using the parameters of background phase in \eqref{eq:bg}, we prepared three training datasets that will be used in the subsequent numerical experiments. These are shown in Table \ref{tab:datasets}. The purpose of each dataset is described as follows: Set 0 is used to experiment with different neural network architecture in section \ref{sec:3} while Set 1 and Set 2 are used to train two deep neural network models for numerical experiments in section \ref{sec:4}. For each dataset, we choose negative values for $\eta$ to mimic a repulsive background since the attractive case is already taken care of by the pole factor in \eqref{eq:sforbound}. It suffices to use the integer values shown in the second column of Table \ref{tab:datasets} for the purpose of this study. Then, for each $\eta$ we generate $500$ random values of $\Lambda_{bg}$ in the range specified in third column of Table \ref{tab:datasets}. The size of each dataset is determined by the parameters of the pole part. 
	
 The parameters for pole part of bound-virtual in \eqref{eq:sforbound} is generated as follows. For each $\eta$ and $\Lambda_{bg}$ in Table \ref{tab:datasets}, we choose $1,000$ random values of $\gamma$ in the interval $(-0.9\Lambda_{bg}, -10\mbox{ MeV})\cup(10\mbox{ MeV}, 200\mbox{ MeV})$. This choice gives a range of bound state binding energy from $0.106\mbox{ MeV}$ to $42.55\mbox{ MeV}$. We ensure that the range of $\gamma$ is cut so that equal numbers of near-threshold virtual and bound state poles are generated. With the values of  $\eta, \Lambda_{bg}$ and $\gamma$ specified, the S-matrix in \eqref{eq:sforbound} can now be used to calculate the input partial wave $|f(E)|^2$ in \eqref{eq:pwa}. For each input, we assign an output label based on the sign of $\gamma$, i.e. label $0$ if $\gamma>0$ (bound state) and $1$ if $\gamma<0$ (virtual state). The number of parameters used results into  a total of $4\times 500\times 1000 = 2,000,000$ input-output samples for bound and virtual state. 	
	
\begin{table}
	\centering
	\caption{
		Dataset generated in this study}
	\begin{tabular}{cccc}
	\hline
	\textbf{Dataset}&
	$\eta$&        	
	$\Lambda_{bg}$ in (MeV)&
	\textbf{Size}\\
	\hline
	Set 0			&
	$[-4,-3,-2,-1]$&
	$(100, 1100)$&
	$2\times 10^6$\\
	Set 1			&
	$[-4,-3,-2,-1]$&
	$(200, 1200)$&
	$4\times 10^6$\\	
	Set 2			&
	$[-4,-3,-1,0]$&
	$(200, 1200)$&
	$4\times 10^6$\\
	\hline	
	\end{tabular} 
	\label{tab:datasets}
\end{table}

\begin{figure}[h!]
	\centering
	\includegraphics[width=\linewidth]{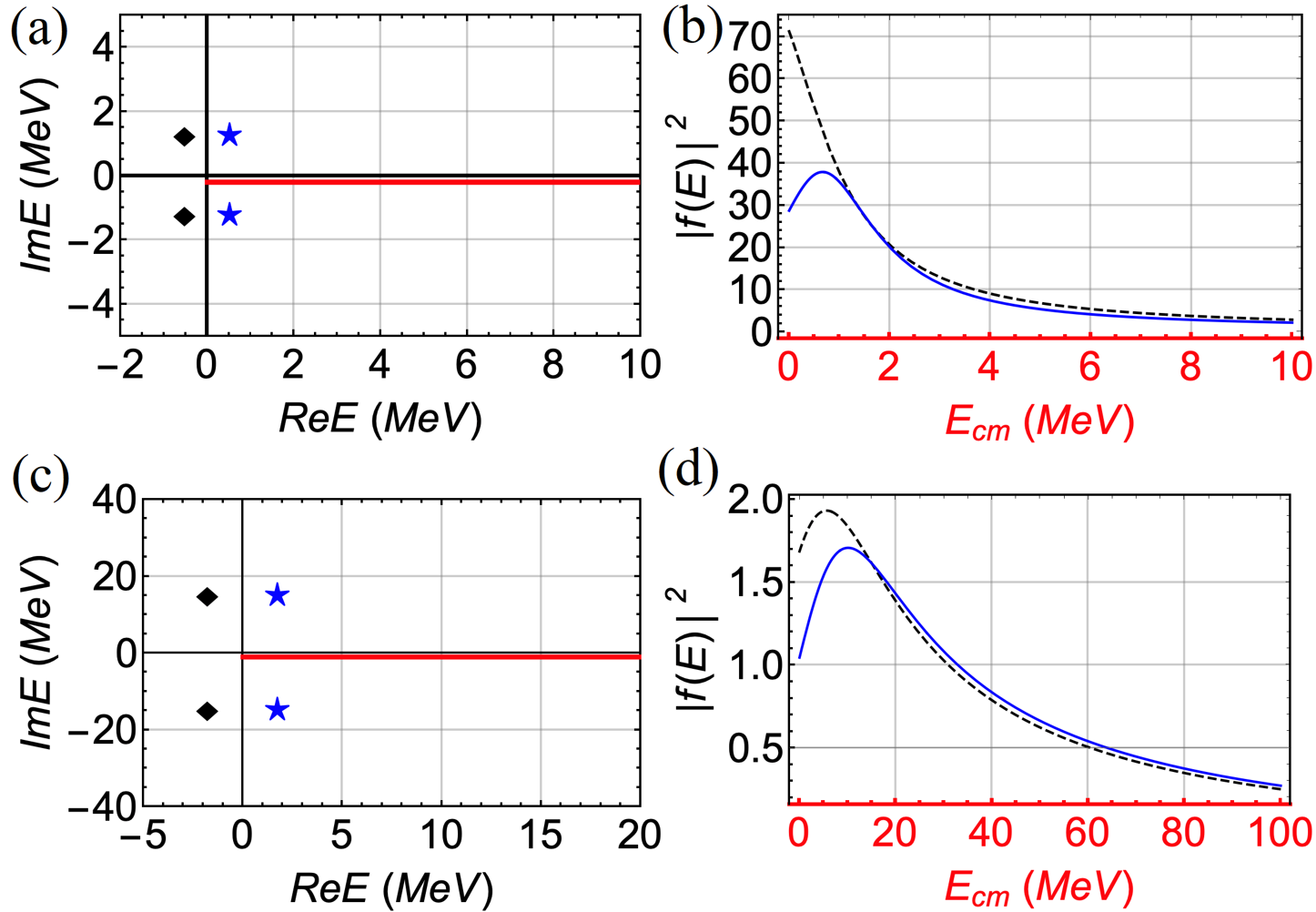}
	\caption{ Virtual ($\diamond$) and resonance ($\star$) poles near threshold (a) and the corresponding line-shape (b). Poles far from the threshold but close to the imaginary axis of unphysical sheet (c) and the corresponding line-shape (d).}
	\label{fig:virtualpeak}
\end{figure}	   	
		
\subsection{virtual state and resonance}
Using the same background phase in \eqref{eq:bg} and the second factor of \eqref{eq:factored}, the S-matrix with conjugate pair of poles is written as
\begin{equation}
	S(p)=e^{2i\delta_{bg}(p)}
	\dfrac{(p-i\beta)^2-\alpha^2}{(p+i\beta)^2-\alpha^2}.
	\label{eq:sforresonance}
\end{equation}
The values of $\eta$ and $\Lambda_{bg}$ are again chosen from Table \ref{tab:datasets} but this time we only choose  $50$ random values for $\Lambda_{bg}$. For the pole parameters, $100$ values of $\beta$ is chosen in the interval $(50\mbox{ MeV}, 200\mbox{ MeV})$ and $100$ values of $\alpha$ in $(1\mbox{ MeV}, 300\mbox{ MeV})$. These choice can give us resonance-peaks with width ranging from $0.12\mbox{ MeV}$ to $64\mbox{ MeV}$. We calculate the input amplitude $|f(E)|^2$ using the above parameters and assign an output label of  $1$ for virtual state pole ($\beta>\alpha$) and $2$ for resonance ($\beta<\alpha$). This is just a continuation of output assignment in the previous subsection. We have a total of $4\times 50\times 100 \times 100=2,000,000$ input-output samples for resonance-virtual classification. 
	
It is interesting to point out that enhancement due to a resonance pole is not completely distinguishable from that of a virtual state pole. Both of these singularities are capable of producing near-threshold peak structures in the scattering region as shown in Fig.\ref{fig:virtualpeak}(d). This is true if we include a background phase in the S-matrix as in \eqref{eq:sforresonance}. A virtual state pole ($\beta>\alpha$) that are far from threshold but close to the imaginary axis of unphysical sheet, as shown in Fig.\ref{fig:virtualpeak}(c), will produce a peak above the threshold due to the distortion caused by the branch point. Normally, if there is no S-matrix background, the conjugate partner of virtual state with width is sufficient to suppress the appearance of peak even if the poles are far from threshold \cite{Doring2009}. This is no longer the case in the presence of background and the conjugate pole must be near the threshold to suppress the peak appearance as demonstrated in Fig.\ref{fig:virtualpeak}(b).
	
A slightly different scenario happens for resonance pole and its conjugate. If it is close to the threshold, a peak structure appears close to the real part of the pole. Here, the conjugate partner is already blocked by the branch cut and can no longer modify the line shape of amplitude. If the resonance pole is moved away from threshold but close to the imaginary axis, the branch point causes the peak structure to appear farther from the pole's real part, resulting to almost identical line shape as that of the virtual pole (see Fig.\ref{fig:virtualpeak}(d)). It is therefore crucial to have a neural network trained to distinguish between these two almost-identical peak structures.
	
\section{Architecture and Training}
\label{sec:3}
Now that we have the classification dataset ready, we proceed with the construction of neural network. To determine the optimal architecture for our task, we experiment with different architectures. Chainer framework \cite{Chainer} is used to build the neural network and to carry out the training. Here, we only use the Set 0 of Table \ref{tab:datasets} which consists only of bound-virtual samples. This dataset is chosen to deliberately make the classification difficult by putting some of the relevant pole in the branch cut of background. We further split the classification data set into two such that $80\%$ is used for training, which optimizes the weights and biases, and the remaining $20\%$ for testing. 
	
Four neural network architectures are used in this experiment. We describe them using the notation
\begin{equation}
	\left[ N_0+1, \cdots, N_{\ell}+1, \cdots, N_{L}+1,3  \right]
	\label{eq:architecture}
\end{equation}
where $N_{\ell}$ is the number of nodes in the $\ell^{th}$ layer ($\ell=0,1,\cdots,L$), with $L$ as the total number of hidden layers and $(+1)$ denotes the added bias. For all architectures, we have $N_0=200$ nodes for the input layer and three nodes for the output.   We assign the ReLU as activation function for hidden-layer nodes
\begin{equation}
	\mbox{ReLU}\left(z_i^{(L+1)}\right)=\mbox{max}\left(0,z_i^{(L+1)}\right)
	\label{eq:relu}
\end{equation}
and use softmax for output nodes
\begin{equation}
	\mbox{softmax}\left(z_i^{(L+1)}\right)=
	\dfrac{\exp\left(z_i^{(L+1)}\right)}
	{\Sigma_{j}^{N_{L+1}}\exp\left(z_j^{(L+1)}\right)}.
	\label{eq:softmax}
\end{equation}
In the classification problem, the cost-function to be minimized is the softmax cross entropy given by
\begin{equation}
	C(\hat{w},\vec{b})=\dfrac{1}{X}\sum_{\vec{x}}\vec{a}(\vec{x})\cdot
	\log\left[\vec{y}_{\hat{w},\vec{b}}(\vec{x})\right]
	\label{eq:cost}
\end{equation} 
where $\hat{w}$ is the weight matrix, $\vec{b}$ is the bias vector, $\vec{x}$ is one of the training input with $\vec{a}(\vec{x})$ as the correct answer, $X$ is the size of training sample and $\vec{y}_{w,b}(\vec{x})$ is the network's output. We use the standard stochastic gradient descent \cite{Bottou2016, Ge2015} to optimize the weights and biases with learning rate of $0.01$ and batch size of $1600$.
	
\begin{figure}[t!]
	\centering
	\includegraphics[width=\linewidth]{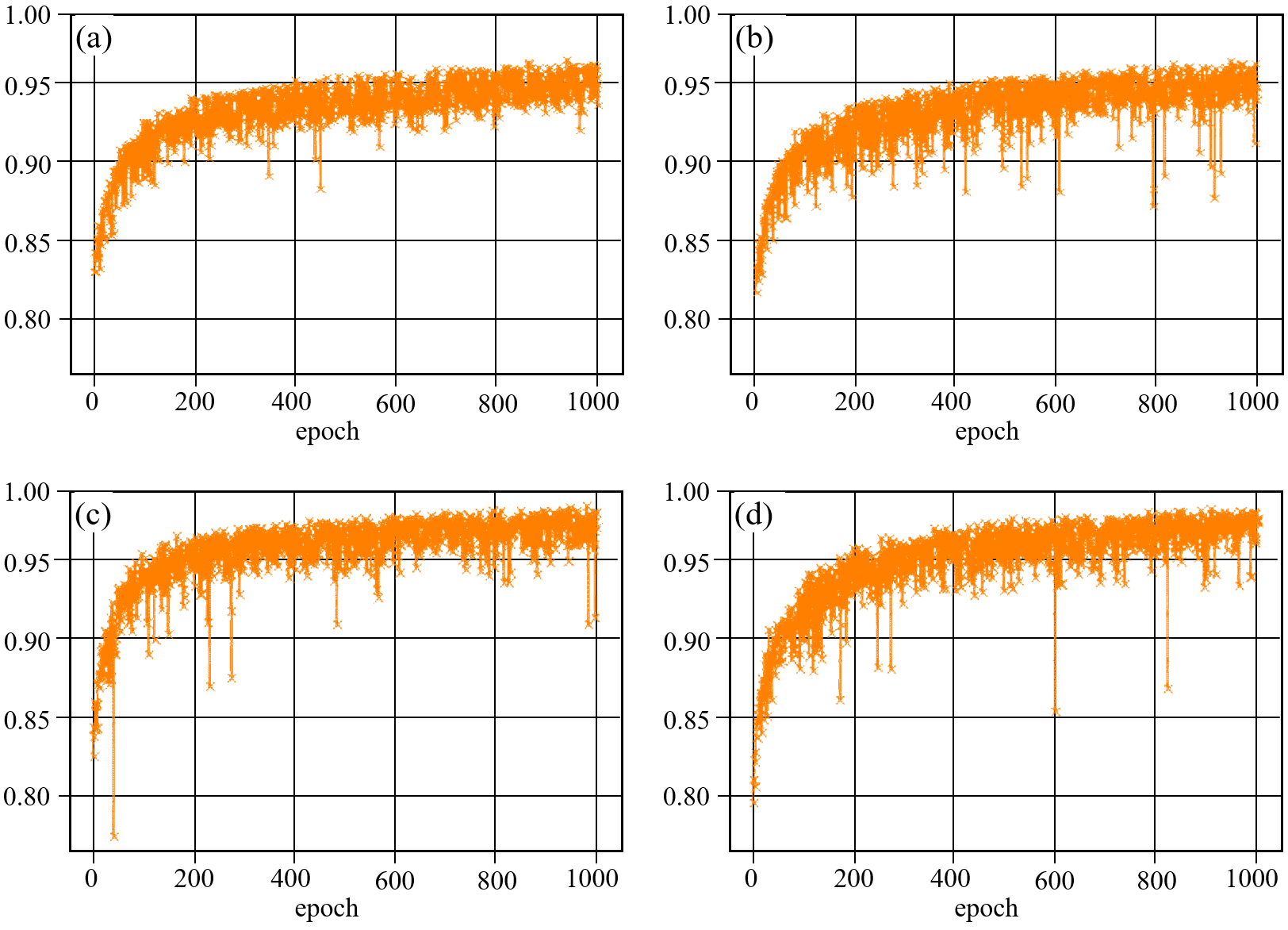}
	\caption{Testing accuracy of different neural networks with architecture 
		(a) $[200+1, 100+1, 3]$, 
		(b) $[200+1, 150+1, 3]$, \\
		(c) $[200+1, 100+1, 50+1, 3]$ and \\
		(d) $[200+1, 50+1, 50+1, 50+1, 3]$}
	\label{fig:archi_search}
\end{figure}	   	
	
The performance of each network architecture is measured by feeding the testing input to the network and comparing the network's output to the correct label. Then, we count the number of correct predictions. The test accuracy of each architecture is shown in Fig.\ref{fig:archi_search}. The vertical axis gives the accuracy of neural network's predictions using the testing set and the horizontal axis is the training epoch.  Generally, the testing accuracy shows large fluctuation due to the stochasticity introduced in the calculation of cost-function. It is interesting to find that the performance of $L=1$ architectures shown in Fig.\ref{fig:archi_search}(a) and Fig.\ref{fig:archi_search}(b) did not improve much even if we added more nodes. After 1000 epochs, the testing accuracies are $94.4\%$ for the $N_1=100$ architecture and $94.5\%$ for the $N_1=150$. This is just a $0.1\%$ improvement in accuracy. However, we get a significant increase when the additional $50$ nodes are placed in the second hidden layer. For a deep neural network with $L=2$, $N_1=100$ and $N_2=50$, the performance is shown in Fig.\ref{fig:archi_search}(c). Here, we get a  $97.2\%$ testing accuracy after 1000 epochs, a significant improvement compared to $L=1$ architecture with the same number of nodes. We also check if increasing $L$, while keeping the total number of nodes fixed, will further improve the performance. The result of $L=3$ with $N_1=N_2=N_3=50$ is shown Fig.\ref{fig:archi_search}(d) giving a testing accuracy of $97.3\%$ after 1000 epochs. The result is almost comparable with the $L=2$ architecture. However, the $L=2$ architecture is more practical to use since it is much faster to train compared to $L=3$. Specifically, for the same number of epochs, the total elapsed time of training for the two $L=1$ architectures are $2.0\times10^5\mbox{ sec.}$ and $2.3\times10^5\mbox{ sec.}$, respectively. While for $L=2$ and $L=3$, we have $2.6\times10^5\mbox{ sec.}$ and $3.1\times10^5\mbox{ sec.}$, respectively. Thus, for the rest of this study we will use a two-hidden layer neural network described in Table \ref{tab:archi}.
	
\begin{table}[t!]
		\centering
		\caption{
			Our Deep Neural Network Architecture}
		\begin{tabular}{lcc}
			\hline
			\textbf{Layer} 						 &
			\textbf{Number of nodes}	&        	
			\textbf{Activation Function}  \\
			\hline
			Input 										&
			200+1   										&
			\\
			1st   										&
			100+1   									  &
			ReLU      								 \\
			2nd   									&
			50+1   									 &
			ReLU      							   \\
			Output   								&
			3   										 &
			Softmax      							\\
			\hline							    
		\end{tabular}
		\label{tab:archi}
\end{table}

\begin{figure}[t!]
	\centering
	\includegraphics[width=0.48\linewidth]{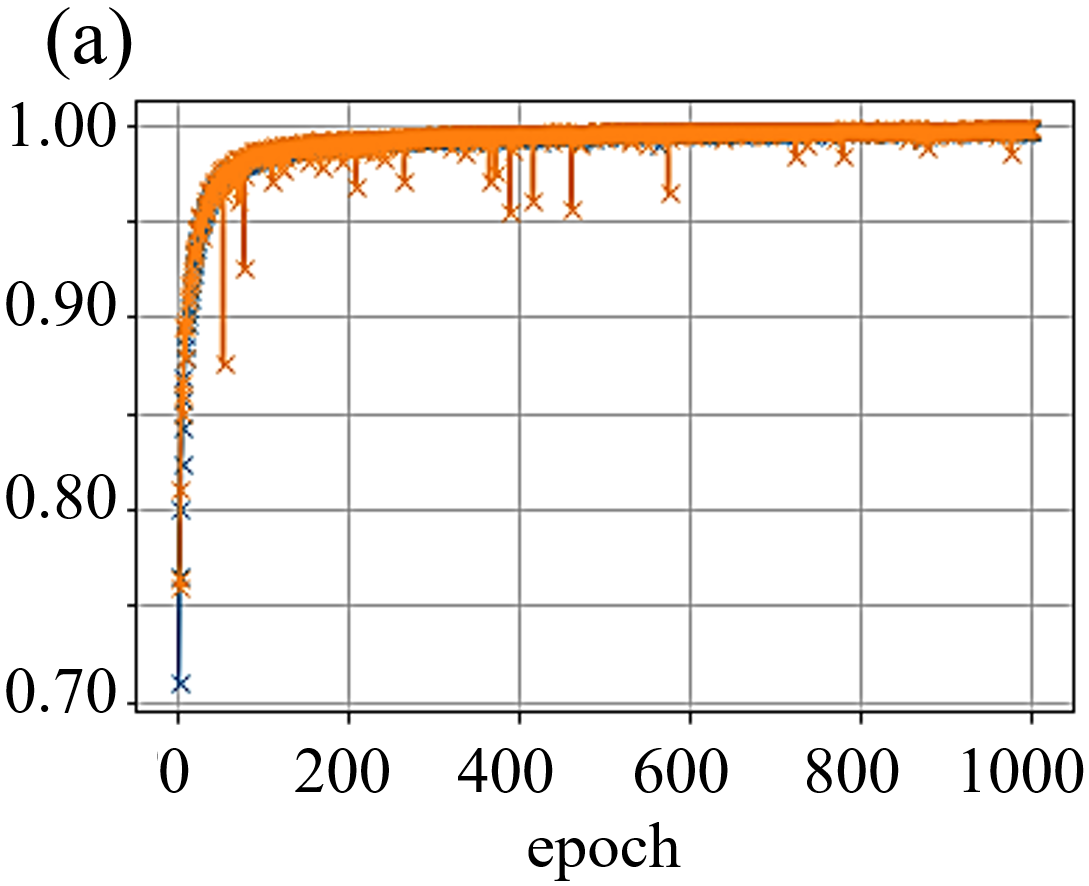}
	\includegraphics[width=0.48\linewidth]{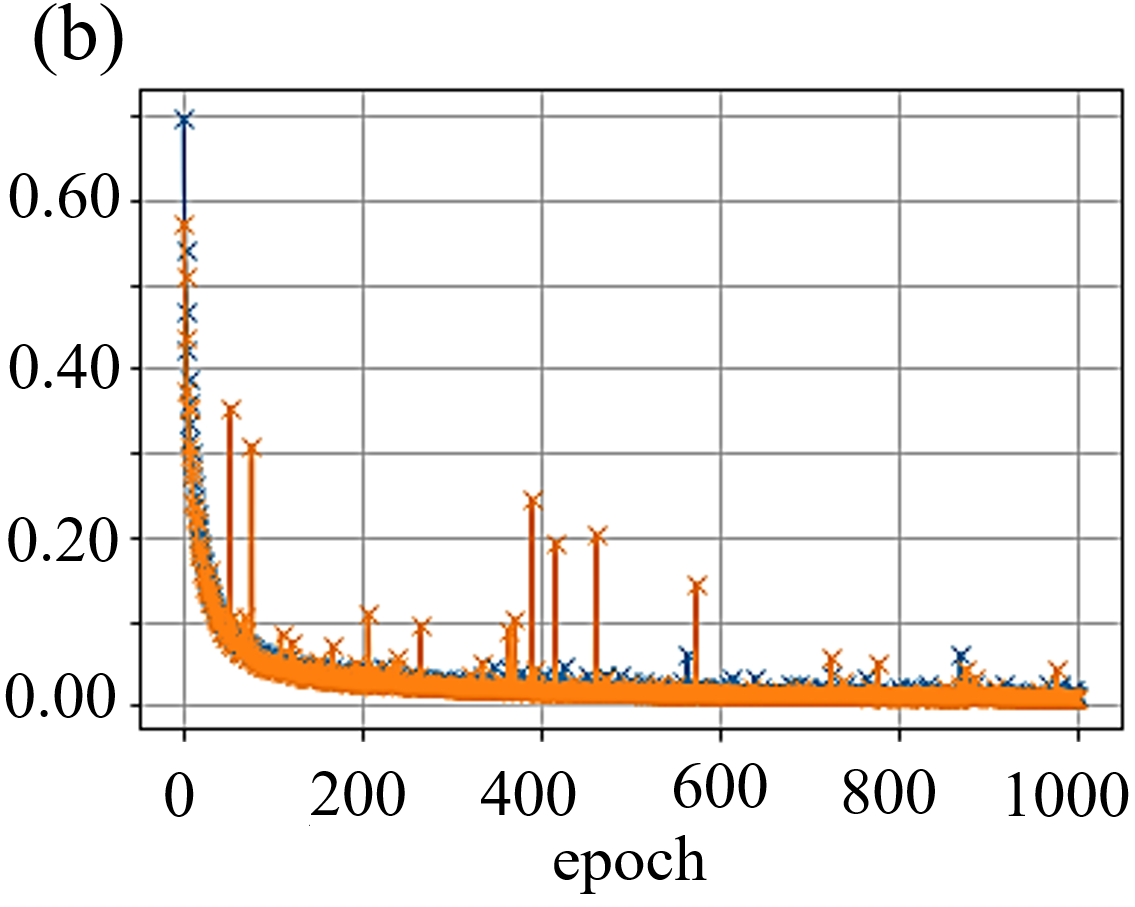}
	\caption{Testing accuracy of neural network model trained using Set 1 (a) and the cost-function profile (b) in each training epoch.}
	\label{fig:threeclass1}
\end{figure}	   	
\begin{figure}[t!]
	\centering
	\includegraphics[width=0.48\linewidth]{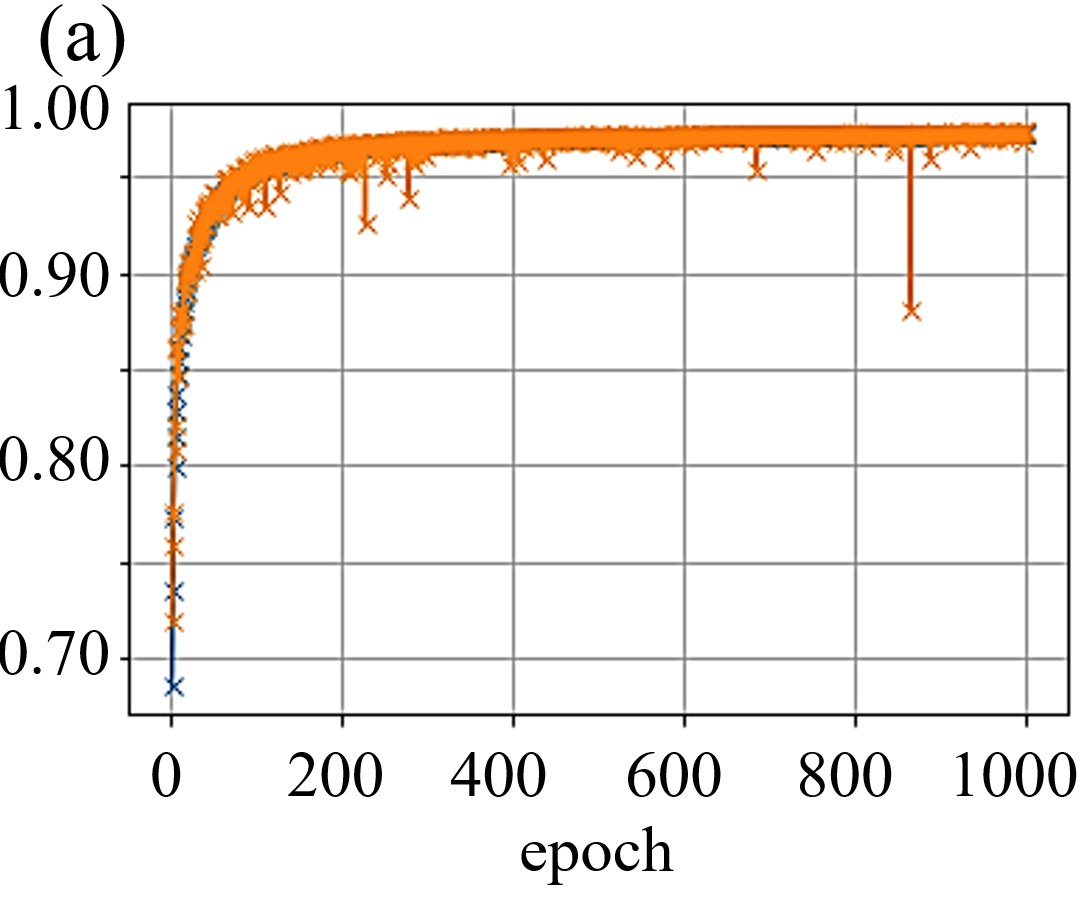}
	\includegraphics[width=0.50\linewidth]{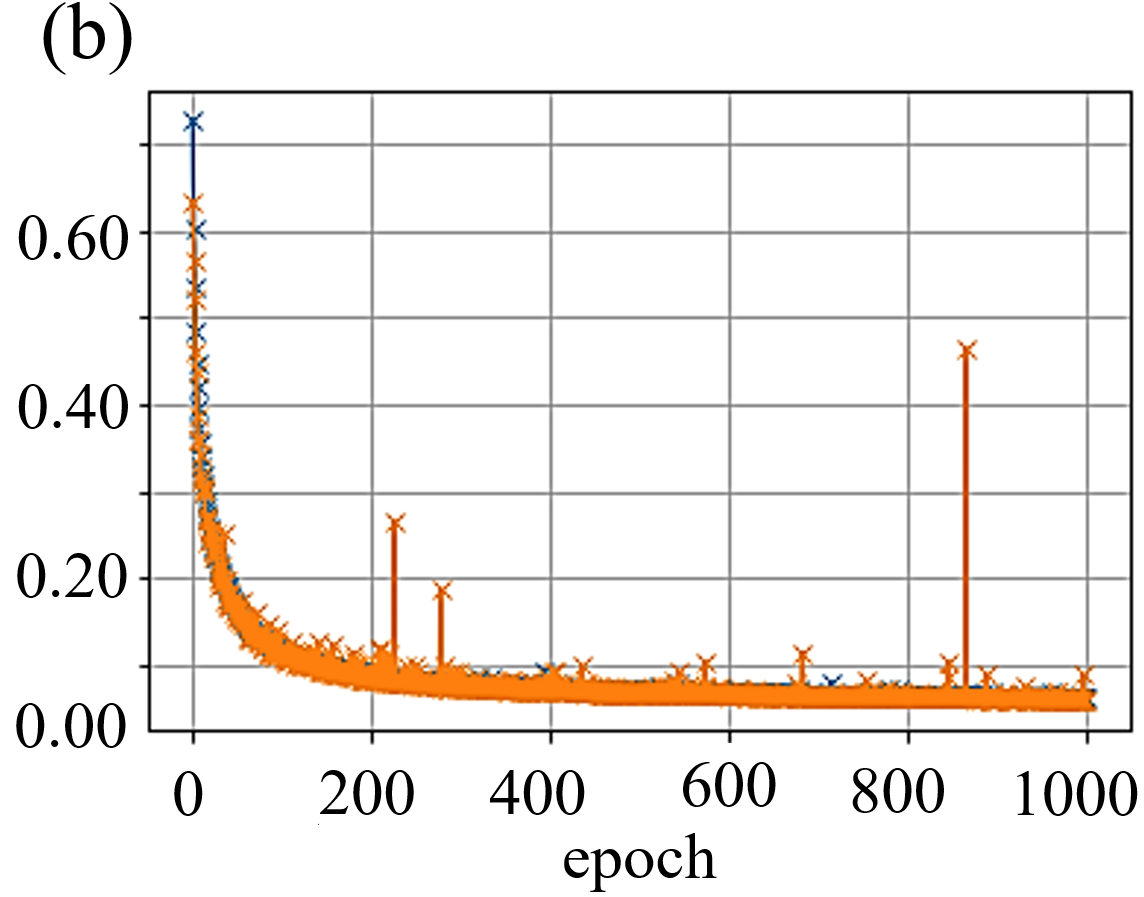}
	\caption{Testing accuracy of neural network model trained using Set 2 (a) and the cost-function profile (b) in each training epoch.}
	\label{fig:threeclass2}
\end{figure}
	
We now proceed to train our chosen network architecture using the classification Set 1 and Set 2 datasets in Table.\ref{tab:datasets}. Each of these dataset contains  $4,000,000$ training input-output tuples for bound-virtual and resonance-virtual cases. The network's performance with Set 1 and Set 2 datasets are shown in Fig.\ref{fig:threeclass1} and Fig.\ref{fig:threeclass2}, respectively. Optimization using Set 1 shows that the accuracy saturates as early as $400$ epochs, indicating that the global minimum of the cost-function is already reached. The network's accuracy is $99.7\%$ for the testing of Set 1 dataset after $1,000$ epochs. The same saturation behavior is observed for Set 2. However, the accuracy after $1,000$ epochs is only $97.3\%$ for testing. The lower accuracy is due to the inclusion of $\eta=0$ which corresponds to no-background case. This gives rise to identical enhancements at threshold whether the pole is a bound or virtual state. Despite its lower accuracy, this dataset is still useful in our subsequent numerical experiment.

\begin{figure*}[t!]
	\centering
	\includegraphics[width=0.8\linewidth]{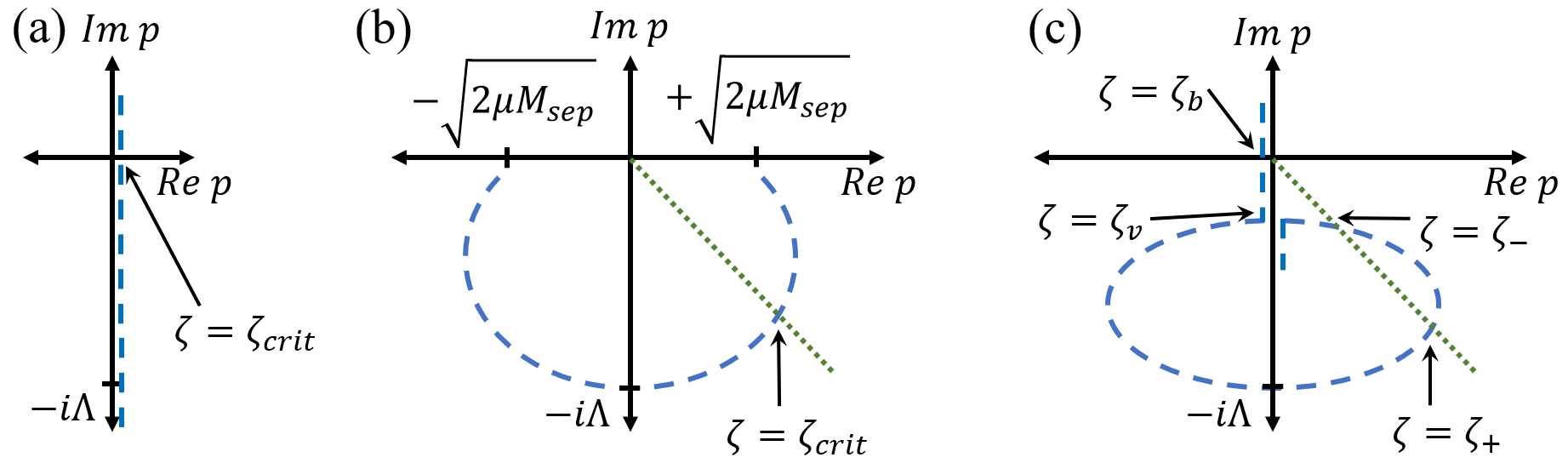}
	\caption{Pole trajectory of separable potential with energy-independent coupling (a), energy-dependent coupling with $M_{sep}>0$ (b) and with $M_{sep}<0$ (c). The dashed line shows the pole's trajectory and the dotted line separates resonances with virtual states.}
	\label{fig:trajectory}
\end{figure*}  

We now have two deep neural network models with the same architecture but trained by two slightly different datasets, i.e. Set 1 and Set 2. In the next section we will study the applicability of these models using an exact solvable separable potential and then apply this to the nucleon-nucleon scattering data.

\section{Validation of Neural Network Model}
\label{sec:4}
We now explore if the trained neural network has the ability to generalize beyond the training dataset. It is important that the validation set be different to that of the training set to make a valid conclusion on the network's ability to generalize. This is done by generating a validation data using an exactly solvable model. 

\subsection{Separable Potential}
The simplest model that can give us an exact solution to the Lippmann-Schwinger equation is a separable potential \cite{Taylor,Newton}. Here, we consider the s-wave potential given by $V(p,p')=\lambda g(p')g(p)$ with Yamaguchi form factor $g(p)=\Lambda^2/(p^2+\Lambda^2)$ where $\lambda$ is an energy-independent coupling strength and $\Lambda$ is a cut-off parameter \cite{Yamaguchi}. The single-channel S-matrix for this model is given by
\begin{equation}
	S(p)=\left(\dfrac{p+i\Lambda}{p-i\Lambda}\right)^2
	\left[\frac{2(p-i\Lambda)^2-\lambda\pi\mu\Lambda^3}
	{2(p+i\Lambda)^2-\lambda\pi\mu\Lambda^3}\right].
	\label{eq:sep_smat}
\end{equation}

We can introduce a dimensionless parameter $\zeta=\pi\mu\lambda\Lambda/2$ to rescale the momentum plane with the cut-off $\Lambda$ as scaling parameter. Fig. \ref{fig:trajectory}(a) shows the trajectory of pole along the imaginary momentum axis as $\zeta$ is varied. At $\zeta=0$, the pole starts at $p=-i\Lambda$ and as $\zeta$ increases in negative value, the pole splits into two. One of the pole moves beyond the cut-off limit while the other one gets closer to threshold. If $-1<\zeta<0$, the near-threshold pole $p_0=i\Lambda(-1+\sqrt{\zeta})$ is a virtual state. If we further make the potential attractive by letting $\zeta<-1$, the near-threshold pole crosses the threshold and becomes a bound state pole. The adjustable parameter $\zeta$ can be used to produce different amplitudes to estimate the network's prediction.	
    
S-wave bound and virtual enhancement at the threshold are possible for separable potential with energy-independent coupling $\lambda$. The absence of centrifugal barrier makes it impossible to produce resonances with attractive interaction \cite{PearceGibson}. This can be modified, however, by allowing the coupling to be energy dependent \cite{Chiral2010}. Minimal number of conjugate poles are produced if we let the energy dependence be
\begin{equation}
	\lambda\rightarrow\left(E-M_{sep}\right)\lambda
	\label{eq:replacelam}
\end{equation}
where $E=p^2/(2\mu)$ with threshold at $E=0$. The parameter $M_{sep}$ is the zero of partial wave amplitude such that when $E=M_{sep}$ there is no scattering. The energy-dependent coupling gives an S-matrix
\begin{equation}
    S(p)=\left(\dfrac{p+i\Lambda}{p-i\Lambda}\right)^2
	\left[
	\frac{2(p-i\Lambda)^2-\lambda\pi\mu\Lambda^3(E-M_{sep})}
	{2(p+i\Lambda)^2-\lambda\pi\mu\Lambda^3(E-M_{sep})}
	\right]
	\label{eq:sep_chiral}
\end{equation}
with the pole position at
\begin{equation}
	\dfrac{p}{\Lambda}=\dfrac{1}{1-\zeta}
	\left[-i\pm\sqrt{\zeta\left(\epsilon\zeta-1-\epsilon\right)}\right]
	\label{eq:chiralpole}
\end{equation}
where we introduce a new set of dimensionless parameters $\zeta=\pi\Lambda^3\lambda/4$ and $\epsilon=2\mu M_{sep}/\Lambda^2$.

Consider the case when the zero of amplitude is on the scattering region, i.e. $M_{sep}>0$ or $\epsilon>0$. We get conjugate pair of poles provided that $\zeta(\epsilon\zeta-1-\epsilon)>0$. This is true for the case of attractive potential, i.e. $\lambda<0$ or $\zeta<0$ and repulsive case when $\zeta>(1+\epsilon)/\epsilon>0$. We consider only the attractive case which is physically meaningful for the discussion of resonance. Fig.\ref{fig:trajectory}(b) shows the trajectory of poles as $\zeta$ is varied. The conjugate poles start at $p=-i\Lambda$ when $\zeta=0$ and moves in the opposite direction as $\zeta$ becomes negative. The pole remains below the line $|Re p|= |Im p|$ when $\zeta>\zeta_{crit}$ where 
\begin{equation}
	\zeta_{crit}=\dfrac{1}{2}\left[\dfrac{1+\epsilon}{\epsilon}
	-\sqrt{\left(\dfrac{1+\epsilon}{\epsilon}\right)^2+\dfrac{4}{\epsilon}}\right]<0.
\end{equation} 	
Here, we only have virtual state with width. If we further make $\zeta$ negative, such that $\zeta<\zeta_{crit}$, the pole will move above the line and turns into a resonance pole. As $\zeta\rightarrow-\infty$, the pole approaches the point $p=\pm\sqrt{2\mu M_{sep}}$ on the real axis. To ensure that the zero will appear in the cross-section, we let the values of $M_{sep}$ to be within $[0,100\mbox{ MeV}]$.
	
The pole trajectory for $M_{sep}<0$ is more involved compared to the previous case. Here, resonance pole can only be produced provided that $-(3-\sqrt{8})<\epsilon<0$, otherwise $\zeta$ will have to be complex. From Fig.\ref{fig:trajectory}(c), we start producing virtual state with widths when $\zeta_{+}<\zeta<0$ and then resonance when $\zeta_{-}<\zeta<\zeta_{+}$ where
\begin{equation}
	\zeta_{\pm}=\dfrac{1}{2}
	\left[\dfrac{1+\epsilon}{\epsilon}\pm
	\sqrt{\left(\dfrac{1+\epsilon}{\epsilon}\right)^2+\dfrac{4}{\epsilon}}\right]<0.
\end{equation}
As $\zeta$ becomes more negative, i.e. $\zeta_{v}<\zeta<\zeta_{-}$ where $\zeta_v=(1+\epsilon)/\epsilon$, the resonance pole will again cross the equal-line and turn into virtual state with width. The two poles will then merge on the zero of amplitude at $p=-i\sqrt{2\mu M_{sep}}$ and then split, producing one near-threshold virtual state pole. This near-threshold pole can turn into a bound state pole if $\zeta<\zeta_{b}$ where
\begin{equation}
	\zeta_{b}=\dfrac{1}{2}
	\left[\dfrac{1+\epsilon}{\epsilon}-
	\sqrt{\left(\dfrac{1+\epsilon}{\epsilon}\right)^2-\dfrac{4}{\epsilon}}\right].
\end{equation}  
\begin{figure*}[t!]
	\centering
	\includegraphics[width=\linewidth]{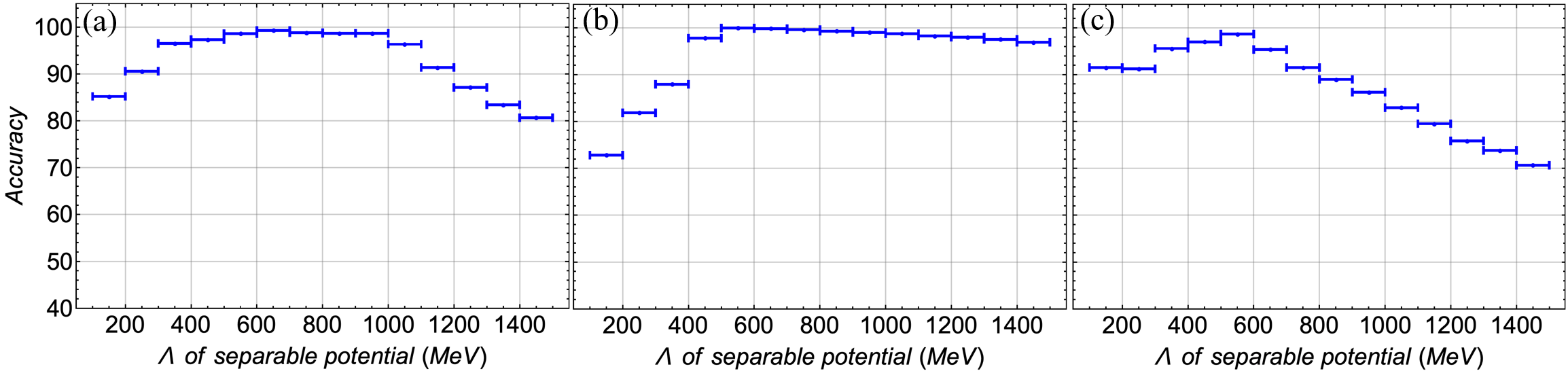}
	\caption{Accuracy of neural network model trained using Set 1 with separable potential amplitude as input for different cut-off parameters. (a) Performance with energy-independent coupling, (b) for energy-dependent coupling with $M_{sep}>0$ and (c) for energy-dependent coupling with $M_{sep}<0$. Each horizontal bar correspond to $100,000$ input s-wave cross-section.}
	\label{fig:set1}
\end{figure*}
	
We separate the validation dataset into three, the first one is generated using the energy-independent coupling which gives amplitude enhancement at threshold. The second and third datasets are generated using the energy-dependent coupling, one with $M_{sep}>0$ and other one with $M_{sep}<0$. The last two datasets are capable of producing peak structures above the threshold. Also, for convenience, we restrict the third dataset, i.e. with $M_{sep}<0$, to produce conjugate poles only. In each set, we choose a range of cut-off parameter $(\Lambda_{min},\Lambda_{max})$ and generated $100,000$ amplitudes using different combinations of parameters.

We must point out that \eqref{eq:sep_smat} and \eqref{eq:sep_chiral} have no background branch cuts along the imaginary axis compared to S-matrix of training data in \eqref{eq:sforbound} and \eqref{eq:sforresonance}. Instead, the validation data has isolated second order pole at $p=i\Lambda$. This might have some repercussions on the predictive power of the trained neural network when applied to the separable potential. 

\begin{figure*}[t!]
	\centering
	\includegraphics[width=\linewidth]{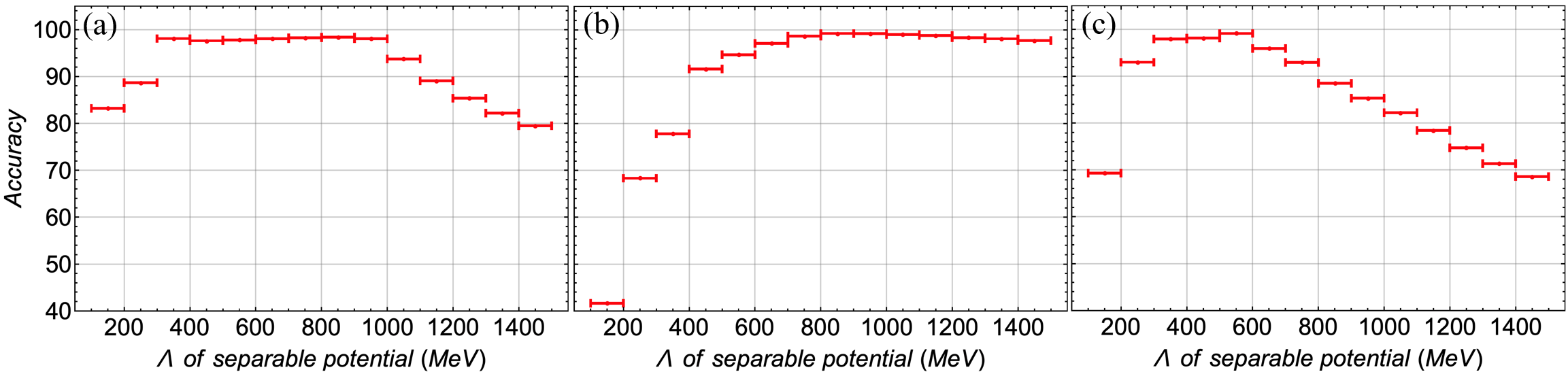}
	\caption{Accuracy of neural network model trained using Set 2 with separable potential amplitude as input for different cut-off parameters. (a) Performance with energy-independent coupling, (b) for energy-dependent coupling with $M_{sep}>0$ and (c) for energy-dependent coupling with $M_{sep}<0$. Each horizontal bar correspond to $100,000$ input s-wave cross-section.}
	\label{fig:set2}
\end{figure*}
	   		
\subsection{Validation of Neural Network Model Trained Using Set 1}	
We now proceed to test our trained neural network using the validation dataset. In particular, we want to investigate if the network can generalize beyond the training set, i.e. we still get accurate predictions even if the validation set is different from the training dataset. Note that if the validation set is just a subset of the training dataset, then we expect that the accuracy of prediction should be high. We also want to explore the region of applicability of the trained neural network. We can asses both the ability of the network to generalize and its applicability by changing the value of cut-off $\Lambda$ since this parameter controls the position of the background singularity. 
	
Consider first the accuracy of prediction with respect to the energy-independent coupling set. From Fig.\ref{fig:set1}(a), we obtain optimal accuracy in the cut-off region between $400$-$1000\mbox{ MeV}$ despite that the background singularity of the validation set is different to that of the training set. We can say that, within this region, the neural network generalizes beyond the training data in distinguishing bound and virtual state enhancements. Below $400\mbox{ MeV}$, the difference between the training and the validation background starts to manifest as seen from the decrease in accuracy as the cut-off is decreased. We also observe a decrease in accuracy in the cut-off region above $1000\mbox{ MeV}$. Here, increasing the cut-off pushes the background far from the scattering region; consequently, a bound or virtual near-threshold pole enhancement becomes identical as we have discussed in section III.
	
It is interesting to find that the accuracy of prediction is different in energy-dependent set as shown in Fig.\ref{fig:set1}(b) and Fig.\ref{fig:set1}(c) even if the neural network is just distinguishing resonance and virtual state with width enhancements for both cases. This difference is probably due to the position of the amplitude’s zero, $M_{sep}$. For the case of $M_{sep}>0$, i.e. the zero is above the threshold, the second order pole background in \eqref{eq:sep_chiral} can produce a bound-like enhancement at the threshold. This is the reason why we get lower accuracy in Fig.\ref{fig:set1}(b) below $400\mbox{ MeV}$. In fact, the network gives a bound state prediction even if there is no bound state in the validation set. This is, however, suppressed in the $M_{sep}<0$ case in Fig.\ref{fig:set1}(c) where the zero below the threshold cancels the effect of the isolated background pole. The absence of extra structure near the threshold allows the network to distinguish a resonance with that of virtual state with width. 
	
The situation is reversed as we go to higher cut-off region. This time, the $M_{sep}>0$ gives high accuracy in $\Lambda>600\mbox{ MeV}$ as shown in Fig.\ref{fig:set1}(b) compared to $M_{sep}<0$ in Fig.\ref{fig:set1}(c). If $\Lambda$ is large, the resonance peak can go beyond the center-of-mass energy range. For $M_{sep}<0$, the zero below the threshold causes the cross-section to monotonically rise from some small value to some maximum at $E_{cm}=100\mbox{ MeV}$. In the absence of peak, the structure for resonance and virtual state with width becomes almost identical. This is the reason why we have decreasing accuracy in Fig.\ref{fig:set1}(c) as the cut-off increases. On the other hand, for $M_{sep}>0$, the large $\Lambda$ means that no bound-like enhancement will appear at the threshold. The structure between the threshold and the zero at $E=M_{sep}$ can still be used by the network to distinguish a resonance with a virtual state with width even if the relevant peak goes beyond the range of center-of-mass energy. This is the reason why we have high accuracy in $M_{sep}>0$ validation set in high $\Lambda$ region.

\subsection{Validation of Neural Network Model Trained Using Set 2}
For certain values of parameters, the training and validation backgrounds can have similar forms. That is, if we set $\eta=-2$, the training background $e^{2i\delta_{bg}}$ reduces to $(p+i\Lambda_{bg})^2/(p-i\Lambda_{bg})^2$ but with domain $\mathbb{C}/(-i\infty,-i\Lambda_{bg})\cup(i\Lambda_{bg},i\infty)$. One may attribute the good performance of our neural network to this similarity. We can test this assumption by using the training Set 2 in Table \ref{tab:datasets} where $\eta=-2$ is replaced with $\eta=0$. The accuracy of the network trained using Set 2 is shown in Fig.\ref{fig:set2}. Notice that above $600\mbox{ MeV}$, the results are all similar to the performance of network trained using Set 1 in Fig.\ref{fig:set1}. This demonstrates that even if the validation dataset is not in the training set, the neural network can still give high accuracy of predictions. This also illustrates that the decrease in accuracy as the cut-off increases as shown in Fig.\ref{fig:set2}(a) and Fig.\ref{fig:set2}(c)  is an intrinsic part of pole classification problem. 
	
We pointed out in the previous subsection that the difference in training and validation background manifests in the low cut-off region. The presence of second order pole in the background of validation dataset and the absence of $\eta=-2$ in the training parameter aggravate the situation. This is seen as a drastic drop in accuracy in Fig.\ref{fig:set2}(b) and Fig.\ref{fig:set2}(c) below $200\mbox{ MeV}$. This means that in this region, the accuracy of the network’s prediction is sensitive to the nature of background singularity. 
	
We give a short comment on the network’s performance on the shallow bound and virtual state produced by energy-dependent set with $M_{sep}<0$. From the trajectory of poles in Fig.\ref{fig:trajectory}(c), a near-threshold bound state or virtual pole is always accompanied by another virtual pole. The latter pole is much closer to the scattering region compared to the accompanying virtual pole of \eqref{eq:sep_smat} in Fig.\ref{fig:trajectory}(a). This makes the classification difficult, i.e. accuracy is less than $50\%$, because the training S-matrix in \eqref{eq:sforbound} educates the network only with single near-threshold pole. This can be improved by inserting an extra pole factor in \eqref{eq:sforbound} to simulate this background virtual pole.

\begin{table}[t!]
	\centering
	\caption{
		$\chi^2/N$ for the Nijmegen partial wave and 
		potential models in the $0-350\mbox{ MeV}$ laboratory frame energy interval. Data for PWA93, NijmI, NijmII, Nijm93 and Reid93 are taken from \cite{Nijmegen1993} and ESC96 is from\cite{Nijmegen1996ECS}.}
	\begin{tabular}{lcccccc}
		\hline
		{}						
		&\textbf{PWA93}
		&\textbf{ECS96}
		&\textbf{NijmI}
		&\textbf{NijmII}
		&\textbf{Nijm93}
		&\textbf{Reid93}
		\\ \hline
		{$N_{par}$}
		&39
		&14
		&41
		&47
		&15
		&50
		\\
		{$\chi^2/N$}
		&0.99
		&1.26
		&1.03
		&1.03
		&1.87
		&1.03
		\\\hline							    
	\end{tabular}
	\label{tab:nijmegenmodels}
\end{table}

\subsection{Application to Nucleon-Nucleon System}
As a final validation, we use the partial wave analyses and potential models of the Nijmegen group \cite{NijmegeSite,PWA93,Nijmegen1993,Nijmegen1996ECS} as input to our neural network. These models are fitted to the nucleon-nucleon scattering data published between 1955 to 1992. They give the correct phase shifts at any laboratory kinetic energy below $350\mbox{ MeV}$. The fitting results are summarized in Table \ref{tab:nijmegenmodels}.  Here, PWA93 corresponds to the analyses of multienergy partial wave on the $pp$ data, the $np$ data and on the combined $pp$ and $np$ database \cite{PWA93}. All three analyses give an excellent fit of $\chi^2/N\sim1$ where $N$ denotes the number of scattering data. Nijm93 is the Nijmegen soft-core potential model introduced in \cite{Nijmegen1993} with NijmI as the nonlocal Reid-like and NijmII is the local version. In the same paper, Reid93 is also introduced which is a regularized Reid soft-core potential. All of these contain the charge-dependent one-pion exchange tail. Lastly, two meson-exchange is included in the extended soft-core ECS96 model of \cite{Nijmegen1996ECS}. 

Now, using the ${}^1S_0$ and ${}^3S_1$ phase-shifts of the mentioned models, we generate the input amplitude on a center-of-mass energy interval $[0,100\mbox{ MeV}]$. We can say that within the cut-off range from $400\mbox{ MeV}$ to $1,000\mbox{ MeV}$, our neural network model can classify a bound-virtual enhancement with $98\%$ accuracy based on our analysis with separable potential model. The resulting amplitude is then fed to the neural network and the results are shown in Table \ref{tab:nijmegenmodelsnetwork}. All the predictions are correct, i.e., the network was able to identify that the ${}^1S_0$ partial wave threshold enhancement is due to the presence of virtual state pole while that of ${}^3S_1$ is due to a bound state pole. It is interesting to point out that the small differences among the models do not affect the network's prediction. This means that if the input data falls within some error band, the neural network can still give consistent classification.

\begin{table}[ht!]
	\centering
	\caption{
		Neural network's prediction with Nijmegen model's amplitude as input. We get the same result whether we use the network trained using either Set 1 or Set 2.}
	\begin{tabular}{lcccccc}
		\hline
		{}						
		&\textbf{PWA93}
		&\textbf{ECS96}
		&\textbf{NijmI}
		&\textbf{NijmII}
		&\textbf{Nijm93}
		&\textbf{Reid93}
		\\ \hline
		{${}^1S_0$}
		&virtual
		&virtual
		&virtual
		&virtual
		&virtual
		&virtual
		\\
		{${}^3S_1$}
		&bound
		&bound
		&bound
		&bound
		&bound
		&bound
		\\\hline							    
	\end{tabular}
	\label{tab:nijmegenmodelsnetwork}
\end{table}        
    	
\section{Conclusion}
\label{sec:6}
This study set out to demonstrate how deep learning can be applied in classifying the nature of pole causing a cross-section enhancement. The method is straightforward in a sense that we can use a simple S-matrix parametrization to generate all the possible line shape that can emerge in the scattering region. We have shown that our neural network model gives high accuracy of more than $90\%$ in the acceptable range of cut-off parameter ($400-800\mbox{ MeV}$). This suffices to have an accurate prediction on the nucleon-nucleon scattering data. Also, the study shows that a neural network trained using a simple S-matrix parametrization is able to generalize beyond the training set. This is demonstrated when we validated our neural network using separable potential models and the nucleon-nucleon Nijmegen models. However, there are limitations in the applicability of deep learning for enhancement classification. One example is the noticeable decrease in accuracy if the cut-off parameter is too large. For the bound-virtual classification, the effect of background is important to distinguish the two structures. While for virtual-resonance classification, the peak structure tend to appear beyond the center-of-energy range if the cut-off is very large, making the classification difficult. 
	
It is important to extend our approach to coupled-channel case since most of the exotic phenomena are believed to be generated from coupled-channel interactions. Although the current study deals with single-channel scattering, the findings can still be used in coupled-channel analysis. In particular, we found that if the validation cut-off is too small, then the neural network's prediction becomes sensitive to the nature of background singularity. This observation should extend to the coupled-channel case and it is appropriate to explore other possible background parametrization such as the one used in \cite{Manley2003,Manley1992}. This will be done elsewhere.
	
\section*{Acknowledgment}
This study is supported in part by JSPS KAKENHI Grants Number JP17K14287,
and by MEXT as ``Priority Issue on Post-K computer'' (Elucidation of the Fundamental Laws and Evolution of the Universe) and SPIRE (Strategic Program for Innovative Research). AH is supported in part by JSPS KAKENHI No. JP17K05441 (C) and Grants-in-Aid for Scientific Research on Innovative Areas, No. 18H05407, 19H05104. DLBS is supported by the UP OVPAA FRASDP and DOST-PCIEERD postdoctoral research grant.


\end{document}